\documentclass[11pt,twoside]{article}
\usepackage[pdftex]{graphicx}
\usepackage{amsmath}
\usepackage{amssymb}
\usepackage{cite}
\usepackage{lineno}

\begin{document}
\newcommand{\pst}{\hspace*{1.5em}}

\newcommand{\rigmark}{\em Journal of Russian Laser Research}
\newcommand{\lemark}{\em Volume 30, Number 5, 2009}

\newcommand{\be}{\begin{equation}}
\newcommand{\ee}{\end{equation}}
\newcommand{\bm}{\boldmath}
\newcommand{\ds}{\displaystyle}
\newcommand{\bea}{\begin{eqnarray}}
\newcommand{\eea}{\end{eqnarray}}
\newcommand{\ba}{\begin{array}}
\newcommand{\ea}{\end{array}}
\newcommand{\arcsinh}{\mathop{\rm arcsinh}\nolimits}
\newcommand{\arctanh}{\mathop{\rm arctanh}\nolimits}
\newcommand{\bc}{\begin{center}}
\newcommand{\ec}{\end{center}}

\thispagestyle{plain}

\label{sh}


\begin{center} {\Large \bf
\begin{tabular}{c}
MODULATION  OF SUPERNARROW EIT PAIR
\\[-1mm]
 VIA ATOMIC COHERENCE 

\end{tabular}
 } \end{center}

\bigskip

\bigskip

\begin{center} {\bf
Lin Cheng$^{1,3,*}$, Zhiyuan Xiong$^{1}$, Shuaishuai Hou$^{1}$, Yijia Sun$^{1}$,  Chenyu Dong$^{1}$, Fan Wu$^{4}$and Kun Huang$^{1,2}$
}\end{center}

\medskip

\begin{center}
{\it
$^1$State Key Laboratory of Dynamic Testing Technology, North University of China, \\Taiyuan, China, 030051

\smallskip

$^2$Upgrading office of Modern College of Humanities and Sciences of Shanxi Normal University,\\
Linfen, China, 041000

\smallskip
$^3$Department of Physics, University of Ottawa,\\ Ottawa, K1N6N5, Canada
\smallskip

$^4$School of Textile Science and Engineering, Xi'an Polytechnic University, \\Xi'an, China, 710049
\smallskip

$^*$Corresponding author e-mail:kiki.cheng@nuc.edu.cn\\}
\end{center}

\begin{abstract}\noindent
We report the phenomena of electromagnetically induced transparency (EIT) and electromagnetically induced absorption (EIA) using two identical beams in rubidium atomic vapor. The $\Lambda$-type EIT configuration is employed to examine the EIT spectrum for the D1 line in $^{87}$Rb F=2 characteristics by varying parameters such as frequency detuning, $I_{\rm probe}/I_{\rm pump}$, the total power of probe and pump beam. Notably, the pump beam is also investigated in this process, which has not been previously studied. We study the effect of of the phase between the two applied fields and find that EIA and EIT can transform into each other by adjusting the relative phase. These finding may have applications in light drag or storage~\cite{1}, optical switching~\cite{2}, and sensing~\cite{3,4}.
\end{abstract}

\medskip

\noindent{\bf Keywords:}
tunable absorber, epsilon-near-zero, metamaterial, bianisotropic response.

\section{Introduction}
\pst
Electromagnetically induced transparency (EIT) enables weak probe light to traverse a medium without absorption under the influence of strong coupling light. Recent years have seen significant theoretical and experimental advancements in the concept and methodology of EIT. EIT has made substantial progress in fields such as nonlinear optics, non-inversion lasers, optical storage, light speed reduction, quantum noise suppression, and quantum information. Notably, advancements have been achieved in quantum noise reduction, nonlinear mixing efficiency, precise measurement of weak magnetic fields, realization of non-inversion lasers, and achieving high refraction with low absorption. The nonlinear phase shift in EIT, associated with light-induced refractive index modulation, can reshape the wavefront of an optical field during propagation and has been observed in various nonlinear media, including liquid crystals, semiconductors, and carbon nanotubes. In addition, the alkali atom systems have recently emerged as powerful platforms for EIT. The multi-parameter can be adjusted, the phenomenon is simple, isotropic, and can be used to verify some basic physical phenomena in Alkali atom which has potential applications in optical switching, high-precision spectroscopy, atomic clocks~\cite{5,6}, and systems with large nonlinearities. It also provides a foundation for practical applications in optical storage~\cite{7} and light drag~\cite{8}. 

Many studies have focused on degenerate EIT using two beams from two lasers. A drawback of this approach is that the overlapping and diameters of the beams are different, necessitating the use of two beams or an acousto-optic modulator. Few studies have examined the use of a single beam emitted by one laser to investigate EIT. Considering that a single beam can form two circularly polarized beams, after a linearly polarized light passes through a 1/4 wave plate, the two beams interact with the medium, avoiding issues of angle and radius inconsistency. This approach is advantageous for studying the EIT effect in degenerate energy levels and can be applied to optical storage applications. This study effectively examines the impact of multiple parameters on EIT under the ideal condition of two beams with the same radius and perfect overlap at an angle of 0 degrees. It provides valuable reference data for degenerate EIT experiments. 

In our experiments, we investigate and analyze the EIT induced by a single-frequency beam, which can be controlled by parameters such as  frequency detuning, ratio between two beams, and total power of these two beams. Meanwhile, the bright and dark states result from constructive and destructive interference between continuous and discrete states, which can be controlled through phase transitions, nonlinear phase shifts, and cross-dressing effects. Our findings are interpreted on the basis of a configuration of Fano interference, utilizing both first- and third-order nonlinearity interference. We also investigated the EIT of pump beam by changing the nonlinear phase while scanning the magnetic field. 

\section{ Experimental scheme}
\begin{figure}[h!]
\centering\includegraphics[width=13cm]{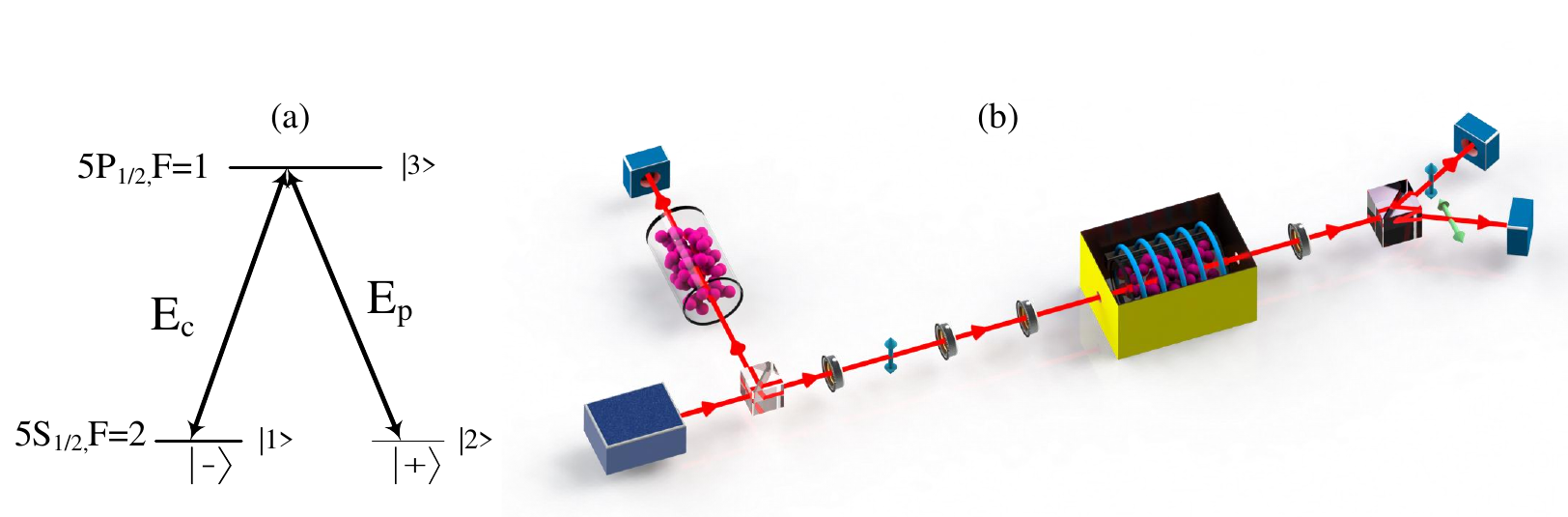}
\caption{(a) $\Lambda$-type configuration of $^{87}$Rb atomic system. The magnetic field B shifts $m=\pm1$ levels by $\pm \delta$. (b) Schematic of the experimental setup.}
\label{Fig1}
\end{figure}  
As shown in Fig.\ref{Fig1} (a), the Rb atoms act as a two-level atomic system, including hyper-fine state-degenerated-Zeeman levels 5$S_{1/2}$, $F=2$ $(|1\rangle$ and $|2\rangle$) and an excited state $5P_{1/2}$ $(|3\rangle)$. The experimental setup is shown schematically in Fig.\ref{Fig1}(b). A collimated Gaussian beam profile with a diameter of 3.2 mm is obtained by expanding the beam from an external cavity diode laser (ECDL) at a center wavelength of $\sim$ 794.88 nm. The Gaussian beam is separated into two arms, the first beam passes through the  Rb cell without buffer gas as a reference beam, and the other beam is seeded into a 7 cm-long Rb cell at the same temperature which  is maintained at 75$^\circ$C by a heated tape. The second beam is decomposed into two circularly polarized sub-beams by a fixed polarizer and rotated HWP denoted as $\bf {E}_p$ (left circularly polarized $\sigma^+$) and $\bf {E}_c$ (right circularly polarized $\sigma^-$), causes the transition $\Delta m = \pm 1$. The outputs of two beam are automatically aligned and propagate perfectly through a magnetically shielded Rb cell. The first beam are seeded into a 7 cm-long thermal Rb vapor cell installed into a solenoid coil and wrapped by $\mu$-metal for magnetic shielding. 
\begin{figure}[htb!]
\centering\includegraphics[width=8cm]{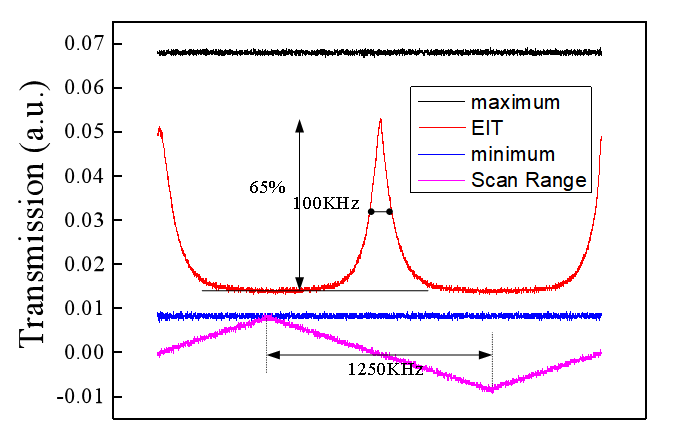}
\caption{ The induced typical EIT signal, the full width of the resonance is about 100 KHz. The EIT signal generated can be up to 65\% efficient.}
\label{Fig2}
\end{figure}
The second Rb cell is filled with about 5 torr of He buffer gas to ensure the long lifetimes of the atomic Zeeman coherence, obtaining narrow EIT resonances. A longitudinal magnetic field for shifting the Zeeman sub-levels is created by a solenoid mounted inside the inner magnetic shield. The cell and the solenoid are enclosed in a $\mu$-metal cylinder to eliminate earth’s and stray magnetic fields. A solenoid is used to control the static magnetic field along the propagation direction of the laser beam. The produced magnetic field addressed a Zeeman coherence tuned to different hyperfine optical transitions with orthogonal linear polarization. The scan range of magnetic field is about 939 mG, corresponding to 1250 KHz. The frequency can be fixed at the resonance point of $^{87}Rb, F=2$, and then the magnetic field is swept, changing the effective detuning.

A wollaston prism is used to separaet the probe and pump (coupling) beams, allowing their transmissions to be monitored as a function of detuning. The transmitted intensities are recorded directly by two photodetectors. The signal obtained from the photodiodes, while scanning the external magnetic field, is recorded by the digital oscilloscope and transferred to a computer. We fix the frequencies of both the frequencies of coupling and probe beams, and change the splitting between (5$S_{1/2}, m_F=+1$) and (5$S_{1/2}, F=2, m_F=-1$) by scanning a bias magnetic field in the same direction as the light propagation. In other words, we scan the detuning of the pump and probe to their respectively resonant transitions. We employ the transition of $|1\rangle\rightarrow|3\rangle$ for probe (horizontally polarized) and $|2\rangle\rightarrow |3\rangle$ for coupling (vertically polarized). The probe and pump beams are perfectly spatial overlapped. The frequency detuning denotes the difference between the resonant transition frequency $\Omega_i$ and the frequency $\omega_i$ of $E_i$ ($i$=1,2). $G_i=\mu_{i3}E_i/\hbar$ is the Rabi frequency between the states $|j\rangle$ and $|3\rangle$ ($j=1,2$), where $\mu_{j3}$ is the transition dipole matrix element. The two beams propagating in the same direction eliminate Doppler effects due to the motion of atoms. We scan the longitudinal magnetic field in this work instead of scanning the probe beam. 
$\gamma_1$ and $\gamma_2$ are rates of spontaneous emission into
the respective states $|1\rangle$ and $|2\rangle$, whereas $\Gamma_{ij}$ describes the decay rate of coherence between $|i\rangle$ and $|j\rangle$ states. When a homogeneous magnetic field is applied to the cell along the light axis using a solenoid coil, the energies of $|1\rangle$ and $|2\rangle$ states vary accordingly due to the Zeeman effect. Additionally, the spectrum has a flat background that is different from other high-resolution spectroscopy techniques using a scanning probe beam. We next present how the signal is influenced by different parameters.

\section{Theory analysis}
 The absorption of probe can be expressed by the absorption coefficient $\alpha=\omega_in_0$Im$(\chi)/c$ \cite{9}, where $\omega$ is the angular frequency of the probe beam, $c$ is the speed of light in vacuum, and $\chi(\omega)$ is the electric susceptibility, $n_0$ is the background refractive index. The absorption of the probe is proportional to Im$(\rho_{31})$. The dispersion coefficient is given by $\beta=\omega_in_0$Re$(\chi)/2c$. To determine the susceptibility $\chi$, we analyze the density matrix $\rho_{ij}$. The on-diagonal density matrix $\rho_{ii}$ correspond to the probabilities of occupying specific states, while the off-diagonal density matrix $\rho_{ij}$ are proportional to the electric dipole moment of the transition and describe the coherence associated with the transition. For the transmission of probe and pump beams, the Liouville (or Von Neumann) pathways are $\rho^{(0)}_{11}\xrightarrow{G_2} \rho^{(1)}_{31}$ (dressed by $E_2$) and $\rho^{(0)}_{22}\xrightarrow{G_1} \rho^{(1)}_{32}$ (dressed by $E_1$), respectively. The expression of the first-order density matrix expanded using Taylor expansion are given by
\begin{equation}
\begin{split}
    \rho^{(1)}_{32}=&\frac{iG_p}{\Gamma_{32}+i\Delta_p+\frac{G_c^2e^{-iv_ct}e^{i\Phi_s}e^{i\Phi_x}}{\Gamma_{31}+i(\Delta_p-\Delta_c)}}\\
     \approx &\frac{iG_p}{\Gamma_{32}+i\Delta_p}\times (1-\frac{G_c^2e^{-iv_ct}e^{i\Phi_s}e^{i\Phi_x}}{(\Gamma_{31}+i(\Delta_p-\Delta_c))(\Gamma_{32}+i\Delta_p)})\\
          =&\frac{iG_p}{\Gamma_{32}+i\Delta_p} +\frac{-iG_pG_c^2e^{-iv_ct}e^{i\Phi_s}e^{i\Phi_x}}{(\Gamma_{32}+i\Delta_p)(\Gamma_{31}+i(\Delta_p-\Delta_c))(\Gamma_{32}+i\Delta_p)}\\
          =&\rho_p^{(1)}+\rho_p^{(3)}
\end{split}
\end{equation}
	\begin{equation}
\begin{split}
    \rho^{(1)}_{31}=&\frac{iG_c}{\Gamma_{31}+i\Delta_c+\frac{G_p^2e^{-iv_pt}e^{i\Phi_s}e^{i\Phi_x}}{\Gamma_{32}+i(\Delta_c-\Delta_p)}}\\
     \approx &\frac{iG_c}{\Gamma_{31}+i\Delta_c}\times (1-\frac{G_p^2e^{-iv_pt}e^{i\Phi_s}e^{i\Phi_x}}{(\Gamma_{32}+i(\Delta_c-\Delta_p))(\Gamma_{31}+i\Delta_c)})\\
     =&\frac{iG_c}{\Gamma_{32}+i\Delta_c}+\frac{-iG_cG_p^2e^{-iv_pt}e^{i\Phi_s}e^{i\Phi_x}}{(\Gamma_{31}+i\Delta_c)(\Gamma_{32}+i(\Delta_c-\Delta_p))(\Gamma_{31}+i\Delta_c)}\\
      =&\rho_c^{(1)}+\rho_c^{(3)}
\end{split}
\end{equation}
where $\delta=g\mu_BB/\hbar$ is the Zeeman level shift of hyperfine level degeneracy caused by an external magnetic field $B$. Here, $\mu_B$ is the Bohr magneton, $g$ is the gyro-magnetic factor, $B$ is the applied magnetic field, $\Phi_0={-v_pt}$ is the initial phase, $\Phi_s$ and the $\Phi_x$ are the self-Kerr nonlinear phase and cross-Kerr nonlinear phase, respectively. The Kerr nonlinear coefficient $n_2$ can be expressed as 
\begin{equation}
n_2=\frac{Re {\chi} ^{(3)}}{\epsilon_0cn_1}=n_2^S+n_2^X=Re\frac{N\mu_{12}(\rho_{21}^{S(3)}+\rho_{21}^{X(3)})}{\epsilon_0E_s^2E_p}\frac{1}{\epsilon_0cn_1}.
   \end{equation}
   Here, $\chi^{(3)}$ is the nonlinear susceptibility that induces the nonlinear phase through self- and cross-phase modulation~\cite{10}.
The imaginary parts of $\rho_{31}$ and $\rho_{21}$ proportionally determine the absorption of the probe and pump beams after passing through the Rb cell. 
The matrix element $\rho$ is related to the complex susceptibility by the relation,
\begin{equation}
\begin{split}
     P=&\frac{1}{2}\epsilon_0E_i\left[\chi(\omega_i)e^{-i\omega_it}+c.c.\right]\\
=&-2\hbar gN\rho e^{-i\omega_it}+c.c.
\end{split}
   \end{equation}
where $N$ is the density of atoms and $c.c$ is the complex conjugate. Substitute $\rho_{ij}$ into Eq. (3) allows the calculation of the absorption and dispersion of the Rb.
The dressed first-order density matrix $\rho^{(1)}_{31}$ can be approximated as the sum of the first order and third-order terms governed by the dressed phase and $\Delta_c$. The interference $|\rho_{11}^{(1)}+\rho_{31}^{(3)}|\approx \sqrt{I_1+I_2+2I_1I_2\cos(\Delta_{\phi})}=|\rho_{11}^{(1)}|\pm|\rho_{31}^{(1)}|$, where $I_1$ and $I_2$ represent the intensities of first order and third order nonlinearity. Therefore, the interference can be constructive or destructive depending on whether the phase to be either 0 or $\pi$. 

According to Eqs.1 and 2, the nonlinear phase shift is influenced by the intensity of the incident beams. To verify this, detailed experiments were set up to study the influence of the parameters on the probe and pump signals.
\section{Results and discussion}
\begin{figure}[htb!]
\centering\includegraphics[width=15cm]{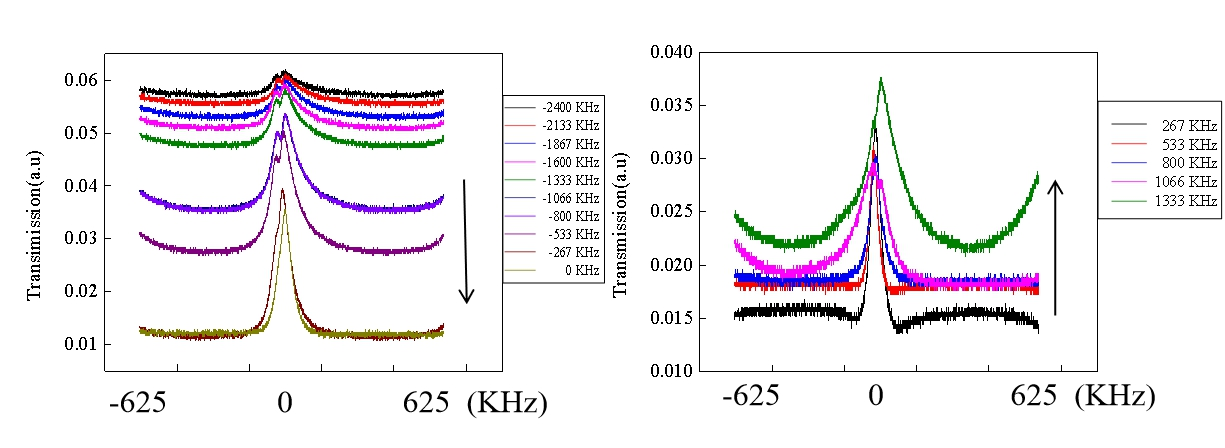}
\caption{Probe transmission as a function of frequency detuning. (a) The frequency detuning increase from negative to 0. (b) The frequency continuously increase from 0 to positive detuning. Change in probe transmission as a function of detuning from $^{87}Rb, F=2$ to $^{85}Rb, F=3$, With a small magnetic field of order 1 G to lift the degeneracy in the $F=1$ and $F=2$ ground states. EIT spectra due to the Zeeman splitting of the magnetic sublevels in a magnetic field.}
\label{Fig3}
\end{figure}
The power of the probe beam ($P_1\approx$ 100 $\rm \mu$W ) and coupling beam ($P_2\approx $ 1 mW) can make the spectral width of the EIT small enough to achieve high efficiency. We applied an external magnetic field to investigate the effects of Zeeman splitting of the magnetic sub-levels. Applying 100 mV to the solenoid generated approximately 626 mG. A magnetic field of 30 mG corresponds to about a 40 KHz level shift. The transmission spectra of the probe beam at the magnetic field of 626 mG are shown in Fig.\ref{Fig3}(a), which illustrates the calculated transmission of the probe beam as a function of its frequency. Here, the magnetic field ranges from 0 to 939 mG. The efficiency of the EIT components gradually increase to 67$\%$ gradually with an increase in detuning, accompanied by an FWMH of 100 KHz. Meanwhile, Autler-Towns splitting appears. In contrast, the width becomes large with an decrease in the detuning without Autler-Townes splitting. A broaden FWMH of EIT is due to response of decay rate ${\gamma}_{87,b}<\gamma_{85,b}$, $5^2S_{1/2}\rightarrow 5^2P_{1/2}$ < $5^2S_{1/2}\rightarrow 5^2P_{3/2}$ and an increase in the magnetic field leads to an increase in the detuning $\Delta$ of the frequency $\omega_c$ from the respective atomic transition. The pure dressing suppression dip corresponds to the dressed states $|\pm\rangle$ created by $E_2$. Then the $E_1$ field only resonates with the dressed energy level $|G_{1+}\rangle$, which meets the enhancement condition $\Delta_1+\lambda_{+}=0$ and a bright state (pure enhancement peak) is observed as shown in Fig.\ref{Fig3}(a). When $\Delta_1=\Delta_2=0$ in Eqs. (1) and (2), Im($\rho_{31}^{(1)}$) and Im($\rho_{21}^{(1)}$) are approximately 0 which means the Rb is transparent to both the probe beam and coupling beam. 

\begin{figure}[ht!]
\centering\includegraphics[width=10cm]{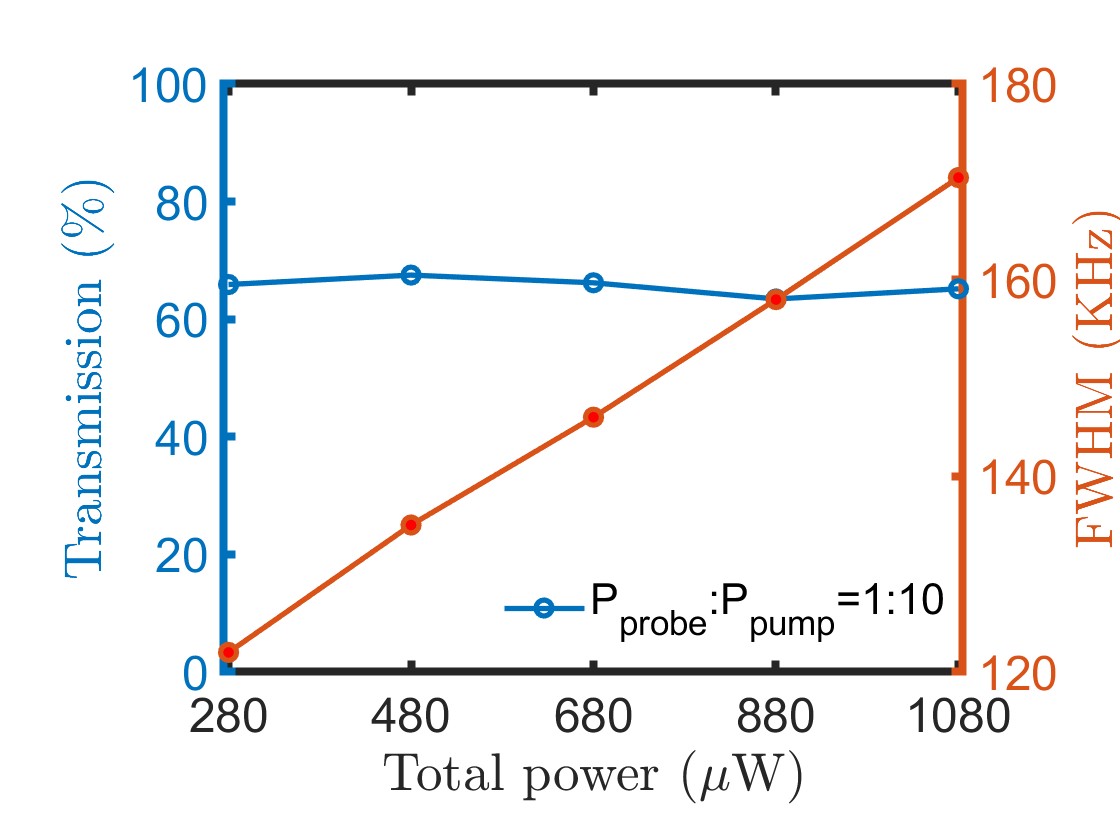}          
\caption{The EIT efficiency and FWMH in different total power with same ratio between probe and pump beam.}
\label{Fig4}
\end{figure}

Furthermore, when we increase the total power while maintaining the ratio between the two beams, the intensity of the EIT signal increases. The experimental results indicate how the intensity of the EIT signal depends on the ratio between the two beam, as shown in the blue line in Fig. \ref{Fig4}. Moreover, Fig. \ref{Fig4} shows that the half-height width in red line increases linearly with the overall power.

\begin{figure}[ht!]
\centering\includegraphics[width=10cm]{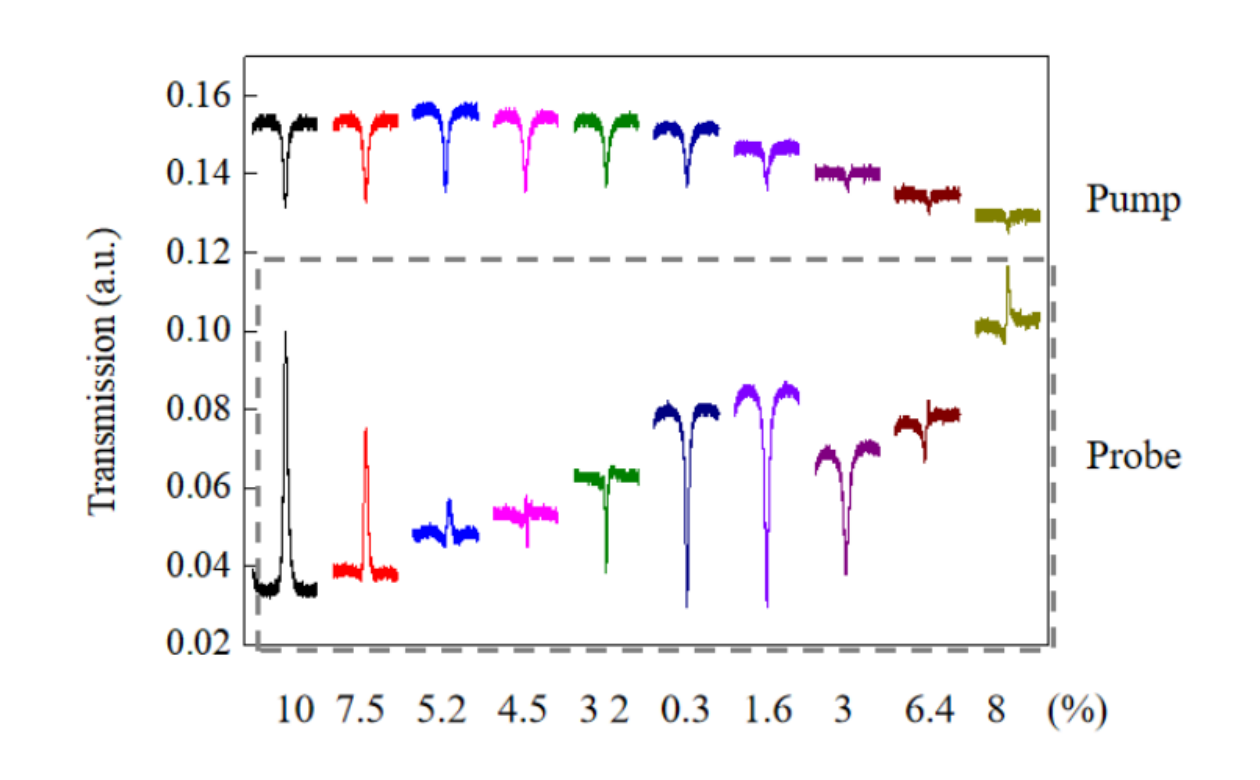}
\caption{Variation of EIT signal with respect to the ratio between probe and pump power, showcasing EIT and EIA conversion. }
\label{Fig5}
\end{figure}

The transmission spectrum of the probe and pump lasers. These two beams exhibit distinct signals, EIT and EIA, and can concurrently measure different group velocities. This optical switch differs from those referenced previously. Changes in probe transmission are shown as a function of detuning from the $F=1-F=2$ levels, with a small magnetic field (approximately 1 G) applied to resolve degeneracy in the $F=1$ and $F=2$ ground states. The range of the transmitted pump light is scaled by a factor of 30. The spectrum makes the system quite sensitive to the relative phase between the two applied fields. The density matrix is always accompanied by a phase-dependent term $\exp(\pm i\Phi)$ where $\Phi$ denotes the relative phase between the two laser fields. Adopting the same parameters as those in Fig. \ref{Fig4}, we obtain two energy absorption curves from the probe and coupling beam as function of the magnetic field for different relative phases. From this figure, we find that under optimal ratio, when $\Phi$ is $\pi$, the system exhibits good EIT, while for $\Phi$ is 0 or $2\pi$, EIA appears. This behavior acts like an optical switch. The FWMH of the observed EIT and EIA FWMH are ~100 KHz. By tuning the relative phase between the two applied fields from $\pi$ to 0, the medium transitions from EIT to EIA, vice versa. The optical Fano interference could be achieved by adjusting experimental parameters. The bright- and dark-state suggests EIT ({I$\propto \lvert{\rho_p^{(1)}}\rvert+\lvert {\rho_p^{(3)}}\rvert$) and EIA (I$\propto \lvert{\rho_p^{(1)}}\rvert-\lvert {\rho_p^{(3)}}\rvert$), respectively. The dressing enhancement peak arises from constructive interference between a perfect continuous state ($\rho_{21}^{(1)}$) and sharp discrete state as shown in Eq. (1) and (2).

\section{Conclusion}
In summary, a technique utilizing a single beam was employed to observe narrow EIT resonances in nearly degenerate $\Lambda$ system rubidium vapor. EIT can be controlled by frequency detuning, ratio between the two beams and total power. Meanwhile, under optimal the relative phase conditions between the two applied fields, EIT rather than EIA can occur at resonance. By tuning the relative phase from 0 to $\pi$, the medium transitions from EIT to EIA and vice versa. The narrow transparency resonances of EIT find applications in magnetometer atomic clocks~\cite{5,6}, light drag~\cite{8}, high-resolution spectroscopy, optic switch and light storage~\cite{7}. 
This approach is expected to significantly enhance the sensitivity of nonlinear magneto-optical measurements substantially. Consequently, we anticipate that this method will attract interest for sensitive optical magnetometer and for establishing new, lower bounds to test the violation of parity and time-reversal invariance. Moremore, as demonstrated here, this technique enables sensitive measurements of phase shifts in nonlinear media with a very high signal-to noise ratio~\cite{11}.

\textbf{Ethical Approval}:not applicable

\textbf{Confilicts of Interest}: The authors declare no confilct of interest.

\textbf{Declarations}
 The statements, opinions and data contained in all publications are solely those of the individual
author(s) and contributor(s) and not of MDPI and/or the editor(s). MDPI and/or the editor(s) disclaim responsibility for any injury to
people or property resulting from any ideas, methods, instructions or products referred to in the content.

\textbf{Author Contributions:} Conceptualization, F.W and L.C; software,L.C;validation, Z. X., Y.S., S.H., C.D; project administration, L.C.; funding acquisition, L.C. All authors have read and agreed to the published version of the manuscript.

\section{Data Availability Statement }
The data that support the findings of this study are available from the corresponding author, upon reasonable request.
\section{Funding}
The support the National Natural Science Foundation of China (62305312), the Natural Science Foundation of Shanxi Province, China (202203021222021), Research Project Supported by Shanxi Scholarship Council of China(2312700048MZ) and the fellowship of China Postdoctoral Science Foundation (2022M722923). Shanxi Provincial Teaching Reform and Innovation Project (2024YB008). 
\section{Acknowledgement}
{L.C acknowledges the support of China Scholaship Council, Zhiyuan Xiong, Yijia Sun, Shuaishuai Hou, Kaixiang Hou, Yi Chen are grateful to L. C for helpful discussions.  }


\end{document}